\documentclass[11pt,a4paper]{article}

\usepackage{epsfig,amsmath,amssymb,cite}
\def\Om{\Omega}
\def\om{\omega}
\def\lp{\left(}
\def\rp{\right)}
\def\lb{\left[}
\def\rb{\right]}
\def\be{\begin{equation}}
\def\ee{\end{equation}}
\begin{document}

\title{Some aspects of cylindrical solutions in Brans--Dicke gravity} 
\author{Andr\'es Arazi$^1$\thanks{e-mail: arazi@tandar.cnea.gov.ar}, 
Claudio Simeone$^{2,3}$\thanks{e-mail: csimeone@df.uba.ar}\\
{\small $^1$  Departamento de F\'{\i}sica, Comisi\'on Nacional de Energ\'{\i}a At\'omica,}\\ 
{\small Av. del Libertador 8250, 1429 Buenos Aires, Argentina}\\
{\small $^2$ Departamento de F\'{\i}sica, Facultad de Ciencias Exactas y 
Naturales,} \\ 
{\small Universidad de Buenos Aires, Ciudad Universitaria Pab. I, 1428, 
Buenos Aires, Argentina} \\
{\small $^3$ IFIBA, CONICET, Ciudad Universitaria Pab. I, 1428, 
Buenos Aires, Argentina}}

\maketitle
\noindent

\begin{abstract}

Several points regarding cylindrical Brans--Dicke geometries are studied. The issue of particle trajectories for the vacuum cylindrical solution is revisited. The possible particular nature of a global string metric is analysed. The general form of the junction conditions for matching cylindrical solutions within this theory is written down, and the relation with the general relativity limit is discussed.\\


\end{abstract}

\section{Introduction}

 In the last decades, cylindrically symmetric geometries have deserved a considerable attention, mainly in relation with cosmic strings. These have been the object of a detailed study, because of the important role they could have played in structure formation in the early universe, and also for their possible observation  by gravitational lensing effects (see Ref. \cite{vilenkin}). Besides,  open or closed fundamental strings are the center of present day attempts towards a unified theory. Then the interest in the gravitational effects of both fundamental and cosmic strings, and in general in axisymmetric solutions, has recently been  renewed  (for example, see Refs. \cite{strings}). 

In Brans--Dicke theory,  matter and non gravitational fields generate a long-range scalar field $\phi$ which, together with them, acts as a source of gravitational field. The metric equations generalizing those of general relativity are 
\be
R_{\mu\nu}-\frac{1}{2}g_{\mu\nu}R=\frac{8\pi}{\phi} T_{\mu\nu}+\frac{\omega}{\phi^2}\phi_{,\mu}\phi_{,\nu}-\frac{\omega} {2\phi^2} g_{\mu\nu}\phi_{,\alpha}\phi^{,\alpha}+\frac{1}{\phi}\phi_{;\mu;\nu}-\frac{1}{\phi}g_{\mu\nu} \phi_{:\alpha}^{;\alpha},\label{ebd1}
\ee 
where $R_{\mu\nu}$ is the Ricci tensor, $T_{\mu\nu}$ is the  energy-momentum tensor of  matter and fields --not including the  Brans--Dicke  field-- and $\omega$ is a dimensionless constant.
The field $\phi$ is a solution of the equation 
\be
\phi_{;\mu}^{;\mu}=\frac{1}{\sqrt{-g}}\frac{\partial}{\partial x^\mu}\left(\sqrt{-g}\ g^{\mu\nu}\frac{\partial\phi}{\partial x^\nu}\right)=\frac{8\pi T}{3+2\omega},\label{ebd2}
\ee
where $T$ is the trace of $T_{\mu\nu}$. With $\phi=\mathrm{constant}=1$ the Einstein's equations are recovered. 

Cylindrically symmetric geometries in Brans--Dicke gravity have been widely studied. Here we shall only deal with some details which we believe deserve a closer look. We shall first revisit the issue of particle trajectories for a general form of the vacuum cylindrical solution. Then we shall analyse the possible particular nature of a global string metric. Finally we shall discuss the general form that junction conditions take for the matching of cylindrical solutions within this theory.

\section{Particle trajectories}

In Ref. \cite{as00b} the weak field approximation of a cylindrically symmetric geometry in Brans--Dicke gravity was analized for the case of an energy-momentum tensor associated to a wiggly cosmic string. It was found that the force on rest --or non relativistic-- particles is attractive, and that no unbounded trajectories exist for massive particles in that geometry as long as the linearized equations remain valid. Besides, under the same approximations, any photon with non vanishing momentum paralell to the string axis is bounded. The natural question is, of course, if that interesting result regarding massive particles and  photons is physically meaningful, or, instead, is merely an artifact yielding from the weak field solution, which is not valid as one goes far away from the string. Here we shall then analize this point by considering the solution to the full theory. We shall identify the case in which the force on massive rest particles points towards the symmetry axis, and then, for the corresponding values of the parameters characterizing the geometry, we shall study the trajectory of both massive and massless particles. 

The general static solution with cylindrical symmetry of the Brans--Dicke equations (\ref{ebd1}) and (\ref{ebd2}) in the vacuum case $T_{\mu\nu}=0$ is given by the metric \cite{as00,bd09}
\begin{equation}
ds^2 = -r^{2d(d-n)+\Om(\om)}\lp dt^2 -dr^2\rp +W_0^2r^{2(n-d)}d\varphi ^2+r^{2d}dz^2.
\label{e11}
\end{equation}
The parameters $n$ and $d$ are constants of integration; $n$ is related with the departure from pure general relativity --see below-- and $d$ can be understood as a mass parameter: for example, in the general relativity limit, $d=0$ corresponds to a solution which is invariant under boosts in the direction of the symmetry axis, and which exerts null force on rest particles.  We have introduced the definition\footnote{In Ref. \cite{as00} there is a typo in Eq. (31), where a factor $(n-1)$ multiplying $\om$ is missing; this can easily be deduced by comparison with the preceding equations.} $\Om(\om)=[\om(n-1)+2n](n-1)$.
With this notation, the scalar field takes the form $\phi=\phi_0r^{1-n}$. The solution for Einstein's gravity is obtained for $n=1$ (which corresponds to a uniform field $\phi=\phi_0$), and is the well known Levi--Civita metric. 

The case of interest for us will be when the parameters $d$ and $n$ are such that a rest particle is attracted towards the axis of symmetry; this corresponds to $\Gamma_{00} ^r > 0$. A simple calculation gives
\be
\Gamma_{00} ^r= \frac{2d(d-n)+\Omega(\om)}{r}
\ee
so that the condition for an attractive behavior is $2d(d-n)+\Omega(\om)>0$. Note that the area per unit length of a cylinder increases with $r$ if $B(r)C(r)=W_0^2r^{2n}$ is an increasing funtion of the radial coordinate, which corresponds to $n>0$, while the length of a circunference increases with $r$ if $n>d$. 

Let us begin by the trajectory of massive particles. Their four-momentum $p_\mu=({\cal E},p_r,p_\varphi,p_z)$ verifies $g^{\mu\nu}p_\mu p_\nu=-m^2$. Then for the metric (\ref{e11}) we have that
\be
-{\cal E}^2+p_r^2+\frac{p_\varphi^2}{W_0^2}r^{2(d-n)+2d(d-n)+\Omega(\om)}+p_z^2 r^{-2d+2d(d-n)+\Omega(\om)}+m^2r^{2d(d-n)+\Omega(\om)}=0
\ee
where ${\cal E}$, $p_\varphi$ and $p_z$ are constants. Defining $L^2=p_\varphi^2 W_0^{-2}$, for the motion restricted to a plane normal to the axis of symmetry we have the ``potential''
$ V^2(r)=L^2r^{2(d-n)+2d(d-n)+\Omega(\om)}+m^2 r^{2d(d-n)+\Omega(\om)}$. Because we are working under the hypothesis that the force on rest particles is atractive, it is $2d(d-n)+\Omega(\om)>0$. Then the second term of the potential for massive particles diverges as $r\to\infty$, and the trajectories of particles of non null mass $m$ are bounded. This result of Ref. \cite{as00b} is then shown to be valid for the full theory, and is not a consequence of the weak field approximation adopted in that work.

Let us now  analyze the behavior of photons. Their four-momentum $k_\mu$ satisfies the identity $g^{\mu\nu}k_\mu k_\nu=0$ so that for the metric considered we have
\be
-k_0^2+k_r^2+\frac{k_\varphi^2}{W_0^2}r^{2(d-n)+2d(d-n)+\Omega(\om)}+k_z^2 r^{-2d+2d(d-n)+\Omega(\om)}=0
\ee
where $k_\varphi$ and $k_z$ are constants. If we define $k_\varphi^2 W_0^{-2}\equiv l^2$ then the behavior of photons in this spacetime is determined by the ``potential'' $U^2(r)=l^2r^{2(d-n)+2d(d-n)+\Omega(\om)}+k_z^2 r^{-2d+2d(d-n)+\Omega(\om)}$. Photon trajectories would be bounded as long as any of the two terms of $U^2(r)$ diverges as $r\to \infty$, and this happens if any of the two powers of $r$ involved is positive. Because we are under the assumption of an attractive force on rest particles, then $2d(d-n)+\Omega(\om)>0$. In particular, in the general relativity limit $n=1$ we have $\Gamma^r_{00}=2d(d-1)/r$ and the force on rest particles is simply determined by $2d(d-1)$, and an atractive force corresponds to $d<0$ or $d>1$. For $n=1$ the potential reduces to $U^2(r)=l^2r^{2(d^2-1)}+k_z^2 r^{2d(d-2)}$. If $d<0$ we have $2d(d-2)>0$, so that the term associated to the momentum along the symmetry axis is divergent; instead, if $d>1$, we have $2(d^2-1)>0$, and the divergent term is the one associated to the angular momentum.  In any case, in the general relativity limit unbounded photon trajectories would not be possible. In the first situation ($d<0$) we obtain the same behavior that in the weak field limit: photon trajectories would be bounded for $k_z\neq 0$. In the Brans--Dicke framework, with $n\neq 1$, the situation is different: for $\Omega(\omega)\leq 0$ the condition of an atractive force on rest particles gives $2d(d-n)>0$, which implies that $d>0$ and $d>n$, or $d<0$ and $d<n$; in the first case the term of the potential proportional to the angular momentum is divergent for $r\to\infty$, and in the second case the term associated to $k_z$ diverges for $r\to\infty$, thus reproducing the result obtained within the weak field limit. But if  $\Omega(\omega)> 0$ then it is easy to verify that the condition of an atractive force is compatible with both terms of the potential decreasing with $r$; bounded photon trajectories are possible, but not unavoidable in this case.

\section{Character of some solutions}

About a decade ago, the geometry associated to a global cosmic string was obtained and studied in detail \cite{sen} in Brans--Dicke gravity. The equations of the theory were solved in the case of cylindrical symmetry, corresponding to the general metric
\be
ds^2=e^{2(K-U)}\lp -dt^2+dr^2\rp +e^{-2U}W^2d\varphi^2+e^{2U}dz^2,
\ee
where $K, U$ and $W$ are functions of $r$, for an energy-momentum tensor of the form $T_t^t=T_r^r=T_z^z=-T_\varphi^\varphi=-v^2e^{2U}/(2W^2)$. The constant $v$ determines the field whose symmetry breaking leads to the existence of the string.  One of the possible solutions is of the form
\be
ds^2=\beta^{-1/m}(r-r_0)^{-1/m}\lp -dt^2+dr^2+dz^2\rp+\beta^{(2\alpha+1)/m}(r-r_0)^{(2\alpha+1)/m}d\varphi^2.\label{sen}
\ee
The constants in this expresion are given by $\beta=\frac{1}{2}v\lp 4\om+7\sqrt{2/(2\om+3)}\rp $, $m=2\om+7/2$ and $\alpha=2\om+2$. If $r_0$ is positive, then the metric is singular at this $r=r_0$. On the other hand,  the author of Ref. \cite{sen} points that  for $r_0=-a<0$ the metric is not singular anywhere, but the limit for $\om\to\infty$ does not reproduce the general relativity solution. Of course, this is true as long as $r$ is positive definite, because if this coordinate is allowed to take negative values then a singularity is present even for $r_0<0$.  But note that in the case that $r\geq 0$, the minimum value $r=0$ does not correspond to the axis of symmetry: for $r=0$ one obtains a circumference of length ${\cal C}=2\pi\beta^{(2\alpha+1)/2m}a^{(2\alpha+1)/2m}>0$, and a cylinder of area per unit of $z$ coordinate ${\cal L}=2\pi\beta^{\alpha/m}a^{\alpha/m}>0$. Moreover, as long as $(2\alpha+1)/2m>0$, the manifold described by metric (\ref{sen}) is such that there is a minimum length curve in a plane normal to the axis, and if, besides, also $\alpha/m>0$, one would have a surface of minimun area per unit of $z$ coordinate at $r=0$.

Let us then recall the notion of a Lorentzian wormhole. A wormhole is a topologically non trivial solution of the equations of gravitation which connects two regions of the universe by a throat.  The definition of the wormhole throat usually adopted for compact configurations characterizes it as a minimal area surface (flare-out condition). In the case of a  cylindrical configuration, the area function ${\cal A}(r)=\sqrt{g_{\varphi\varphi}(r)g_{zz}(r)}$ is defined. Then this flare-out condition implies that ${\cal A}$ should increase at both sides of the throat, and this means  ${\cal A}'>0$, so that $\lp g_{\varphi\varphi}g_{zz} \rp'>0$ (from now on a prime stands for a radial drivative). A different definition of the flare-out condition has been recently  proposed  for infinite cylindrical configurations\cite{brle}. Because a wormhole is defined by its non trivial topology, which is a global property of spacetime, and   for  static cylindrically symmetric geometries the global properties are determined by the behavior of the radius function  ${\cal R}(r)=\sqrt{g_{\varphi\varphi}(r)}$, then the flare-out condition has been defined in Ref.\cite{brle} by demanding that ${\cal R}(r)$ has a minimum at the throat; this implies $g'_{\varphi\varphi}>0$. For a compact wormhole, an unavoidable requirement  within the framework of general relativity is that the supporting matter must be exotic, that is, the energy density $\rho$ and the pressures $p_i$ cannot fulfill  the energy conditions $\rho\geq 0, \rho+p_i\geq 0$. But for non compact wormholes, it was shown that under the less restrictive flare-out condition the energy conditions can be satisfied. 

Any of both definitions for the existence of a throat at $r=0$ connecting two copies of the geometry considered could be fulfilled  if we accept that $r\geq 0 $, and if the parameters take suitable values; an analogous situation would take place for other solutions with the same properties. The definition in terms of the area per unit length requires $\alpha/m>0$, which can only be fulfilled if $\om>-1$ or $\om<-7/4$. But the square root in the definition of $\beta$ imposes the restriction $\om>-3/2$; then $\om>-1$ is the only acceptable choice. In this case we would have a minimum area per unit length at $r=0$, which could be associated to a wormhole-like geometry. On the other hand, the definition in terms of the circumference length would require $\om>-5/4$ or $\om<-7/4$, but, again, one can only admit $\om>-3/2$. Then with $\om>-5/4$ there would exist a miminum length circumference at $r=0$, which could be understood as associated to the throat of a lorentzian wormhole. Note that the energy-momentum from which the construction started satisfies the energy conditions. But the configuration is non compact, and the framework is not the Einstein's general relativity. It has been pointed \cite{star} that in scalar-tensor theories wormholes could avoid exotic matter only if the scalar field is a ghost, which means $\om<-3/2$. However, the proof of such important restriction relies on the compact character of the configurations considered, which would not be the case in the example taken from Ref. \cite{sen}. 

We emphasize that we are only discussing a possible interpretation of the metric considered, and of a class of metrics with analogous features. In fact, if one is interested in the actual construction of a lorentzian wormhole, this could always be achieved starting from any suitable solution (i.e. one satisfying the flare-out condition) by simply applying the well known cut and paste procedure with two copies of the manifold, which yields a matter shell located at the junction surface. This would constitute a particular example of application of the results of the following section.

\section{Junction conditions}

We shall now consider in detail the matching of two metrics at a cylindrical static surface $\Sigma$. The junction conditions in Brans--Dicke theory, which relate the metrics at both sides with the induced energy-momentum tensor on the surface (generalized Darmois--Israel conditions), have the form \cite{suba}
\begin{eqnarray}
 -\langle K^i_j\rangle+ \langle K\rangle\delta_j^i &=&  \frac{8\pi}{\phi_\Sigma}\lp S^i_j-\frac{S}{3+2\om}\delta_j^i\rp,\label{cc}\\
\langle \phi_{,N}\rangle  &=&   \frac{8\pi S}{3+2\omega},\label{pp}
\end{eqnarray}
where $K^i_j$ are the components of the extrinsic curvature, and $\phi_\Sigma$ is the field on the junction surface. The  notation $\langle \cdot \rangle$ stands for the jump across the surface  $\Sigma$, $N$ labels the coordinate  normal to this surface and $S_j^i$ is the  energy-momentum tensor of matter and fields (except the field $\phi$) on the  shell located at $\Sigma$. The normal derivative of the field is given by $\phi_{,N}=\phi_{,a}n^a$, where for a constant radius surface the unit normal vector is $n^a=\eta\lp 0,g_{rr}^{-1/2},0,0\rp$; $\eta=1$ corresponds to a vector pointing outwards, and $\eta=-1$ to a vector pointing to the axis. The quantities $K$ and $S$ are the traces of $K^i_j$ and $S_j^i$ respectively.  The components of the metric and the Brans--Dicke field are continuous across the shell ($ \langle g_{\mu\nu}\rangle=0 $, $\langle \phi\rangle=0 $). Note that in the general relativity limit $\om\to\infty$ the Lanczos equations \cite{lanczos} are recovered.

Let us consider two metrics of the general form 
\begin{equation}
ds^2 = -A_{\pm}(r)dt^2 +B_{\pm}(r)dr^2 +C_{\pm}(r)d\varphi ^2+D_{\pm}(r)dz^2,
\label{e1}
\end{equation}
where $A_{\pm}$, $B_{\pm}$, $C_{\pm}$ and $D_{\pm}$ are positive functions of $r$, and paste them at a cylindrically symmetric surface given by a constant value of the function $C$. The coordinate radius of the shell can be taken as $r=r_-$ on one side, and as $r=r_+$ on the other side; because the coordinates do not need to be same on both sides, one must not necessarily set $r_-=r_+$, and the same stands for the time ($t_-\neq t_+$). The continuity of the scalar field imposes $\phi_-(r_-)=\phi_+(r_+)=\phi_\Sigma$. In order to allow for a general application, we take the vector $n^a$  pointing outwards or to the axis for any of the two metrics joined at the surface $\Sigma$; then we have $\eta_+$ and $\eta_-$ which can both take the values $\pm 1$. 
From Eqs. (\ref{cc}) and (\ref{pp}), a straightforward calculation gives the surface energy density and pressures in terms of the metric coefficients and the scalar field: 
\begin{eqnarray}
\sigma & = & -\ \frac{\eta_+}{8 \pi \sqrt{B_+(r_+)}} \left\{\frac{\phi_\Sigma}{2} \left[
\frac{C'_+(r_+)}{C_+(r_+)} + \frac{D'_+(r_+)}{D_+(r_+)}\right] +\phi'_+(r_+)\right\}\nonumber\\
& &+\ \frac{\eta_-}{8 \pi \sqrt{B_-(r_-)}} \left\{\frac{\phi_\Sigma}{2} \left[
\frac{C'_-(r_-)}{C_-(r_-)} + \frac{D'_-(r_-)}{D_-(r_-)}\right] +\phi'_-(r_-)\right\},
\label{e12}
\end{eqnarray}
\begin{eqnarray}
p_{\varphi}& = &  \frac{\eta_+}{8 \pi \sqrt{B_+(r_+)} } \left\{
 \frac{\phi_\Sigma}{2}\left[\frac{D'_+(r_+)}{D_+(r_+)}+ \frac{B'_+(r_+)}{B_+(r_+)}\right] +\phi'_+(r_+)\right\}\nonumber\\
& & -\ \frac{\eta_-}{8\pi \sqrt{B_-(r_-)} } \left\{
 \frac{\phi_\Sigma}{2}\left[\frac{D'_-(r_-)}{D_-(r_-)}+ \frac{B'_-(r_-)}{B_-(r_-)}\right] +\phi'_-(r_-)\right\},
\label{e13}
\end{eqnarray}
\begin{eqnarray}
p_{z}& = &  \frac{\eta_+}{8 \pi \sqrt{B_+(r_+)}}\left\{ \frac{\phi_\Sigma}{2}
\left[\frac{C'_+(r_+)}{C_+(r_+)} + \frac{B'_+(r_+)}{B_+(r_+)}\right] +\phi'_+(r_+)\right\}\nonumber\\ & & -\ \frac{\eta_-}{8 \pi \sqrt{B_-(r_-)}}\left\{ \frac{\phi_\Sigma}{2}
\left[\frac{C'_-(r_-)}{C_-(r_-)} + \frac{B'_-(r_-)}{B_-(r_-)}\right] +\phi'_-(r_-)\right\}.
\label{e14}
\end{eqnarray}
 The continuity of the metric then requires $B_-(r_-)=B_+(r_+)$, $C_-(r_-)=C_+(r_+)$, $D_-(r_-)=D_+(r_+)$. Besides, in order to satisfy Eq. (\ref{pp}) which relates the trace of the induced energy-momentum tensor and the jump of the normal derivative of the field, $\phi$ and the metric coefficients must fulfill the condition
\begin{eqnarray}
\eta_+\phi'_+(r_+)-\eta_-\phi'_-(r_-) & = & \frac{\phi_\Sigma}{2\om}\eta_+\lb\frac{B'_+(r_+)}{B_+(r_+)}+\frac{C'_+(r_+)}{C_+(r_+)}+\frac{D'_+(r_+)}{D)_(r_+)}\rb\nonumber\\
& &  -\ \frac{\phi_\Sigma}{2\om}\eta_-\lb\frac{B'_-(r_-)}{B_-(r_-)}+\frac{C'_-(r_-)}{C_-(r_-)}+\frac{D'_-(r_-)}{D_-(r_-)}\rb.\label{ppp}
\end{eqnarray}
Equations (\ref{e12})-(\ref{ppp}) give the general form of the junction conditions for two cylindrically symmetric static geometries in Brans--Dicke theory of gravity. For example, the matching of an inner solution $g_{\mu\nu}^-$ with an outer metric $g_{\mu\nu}^+$ corresponds to $\eta_+=\eta_-=1$. Instead, the construction of a thin-shell lorentzian wormhole implies two exterior solutions, so that $\eta_+=1$ and $\eta_-=-1$; if, besides, the two sides of the wormhole are equal, then the equations above simplify considerably. Note that there is no complete freedom in the choice of the joining surface if the parameters are given; once these parameters are fixed, two solutions can be matched only at cylindrical surfaces of certain radius. This is not surprising, as the same result is well known in the case of spherical symmetry (See, for example, the first two articles of Ref. \cite{suba}). Also note that while the Einstein's gravity limit of some cylindrical solutions may only require a particular value of some parameter,  this is not enough at the level of the junction conditions. A simple  example is the vacuum case treated above, for which the general relativity solution corresponds to $n=1$. This value gives $\phi'_\pm=0$ for both solutions matched, so that the left hand side of Eq. (\ref{ppp}) vanishes. However, in general the right hand side will not vanish --because the induced surface energy-momentum tensor is not necessarily traceless (see Eq. (\ref{pp}))-- unless the condition $\om\to\infty$ is also imposed. That is: though the ``bulk'' energy-momentum tensor is traceless --in fact, in this case it vanishes identically, so that the right hand side of Eq. (\ref{ebd2}) is consistent with a uniform field corresponding to the general relativity limit--, the surface energy and pressures are not forced to satisfy $S=-\sigma+p_\varphi+p_z=0$, so that $\om\to\infty$ is necessary to satisfy the junction conditions in the general relativity limit.

\section{Summary}

We have considered some particular aspects of static cylindrical solutions within the context of Brans--Dicke scalar-tensor theory of gravity. First we have studied the motion of particles in a vacuum geometry in order to understand the validity of results that had been found previously in the weak field limit of the theory. Then we have analysed what could be seen as a peculiar behavior of the metric associated to a global cosmic string. Finally, we have written down the general form of the junction conditions for the matching of two static cylindrical metrics in Brans--Dicke gravity, and have briefly discussed the issue of the general relativity limit  in relation with the character of the induced surface energy-momentum tensor.   

\section*{Acknowledgments} This work has been supported by CNEA and CONICET. C. S. thanks E. F. Eiroa for useful discussions.

\end{document}